\documentclass[
aps,
reprint,
superscriptaddress,
amssymb,
amsmath
]{revtex4-1}

\usepackage{graphicx}
\usepackage{dcolumn}
\usepackage{bm}
\usepackage{bm, color}
\usepackage{lineno}
\begin{document}

\title{Strong bulk photovoltaic effect in chiral crystal in the visible spectrum}

\author{Yang Zhang}
\affiliation{Max Planck Institute for Chemical Physics of Solids, 01187 Dresden, Germany}
\author{Fernando de Juan}
\affiliation{Donostia International Physics Center (DIPC), 20018 San Sebastian, Spain}
\affiliation{Ikerbasque Foundation, 48013 Bilbao, Spain}
\author{Adolfo G. Grushin}
\affiliation{Institut Néel,25 rue des Martyrs BP 166, 38042 Grenoble cedex 9}
\author{Claudia Felser}
\affiliation{Max Planck Institute for Chemical Physics of Solids, 01187 Dresden, Germany}
\affiliation{Center for Nanoscale Systems,Faculty of Arts and Sciences, Harvard University,11 Oxford Street, LISE 308 Cambridge, MA 02138, USA}
\author{Yan Sun}
\email{ysun@cpfs.mpg.de}
\affiliation{Max Planck Institute for Chemical Physics of Solids, 01187 Dresden, Germany}

\begin{abstract}
Structurally chiral materials hosting multifold fermions with large topological number
have attracted considerable attention because of their naturally long surface
 Fermi arcs and bulk quantized circular photogalvanic effect (CPGE).
Multifold fermions only appear in metallic states, and therefore, most
studies so far have only focused on the semimetals in compounds with chiral crystal
structures. In this work, we show that the structurally chiral topological trivial
insulators are also exotic states, which is interesting from the application point
of view, owing to their natural advantage to host a large bulk photovoltaic effect
in the visible wavelength region. In the last decades, the shift current in the
visible wavelength region was limited
to be 10 $\mu A/V^{2}$.
By scanning the insulators with chiral structure, we found a class of compounds
with photoconductivity ranging from $\sim20$ to $\sim80$ $\mu A/V^{2}$, which is
approximately one order of magnitude larger than that reported in other real materials.
This work illustrates that the compounds with chiral structure can host both quantum
CPGE and a strong shift current in the second order optical response. Moreover, this
work offers a good platform for the study of the shift current and
its future application by putting the focus on insulator with chiral lattices,
so far overlooked in photovoltaic technologies.

\end{abstract}

\maketitle

Topological semimetals with multifold fermions protected
by chiral crystal structures are becoming a new research frontier
in topological materials. Different from Dirac and Weyl
semimetals, the crystal symmetry protects new quasiparticles with
large Chern numbers and naturally long surface Fermi
arcs~\cite{manes2012existence,bradlyn2016beyond,Niels2018,tang2017multiple,chang2017unconventional}.
Because of the lack of mirror symmetry, a
quantized circular photogalvanic effect (CPGE) was theoretically
proposed~\cite{de2017quantized,flicker2018chiral}.
The new fermions in chiral crystals have attracted significant attention
in the last few years, and the quantized CPGE was experimentally
observed very recently~\cite{Dylan2019}. Since multifold fermions only
appear in metallic states, considerable attention has been focused on
semimetals in this class of compounds, whereas much less attention
has been paid to the insulators. Because the chiral crystal structure
naturally breaks the inversion symmetry strongly, insulators are ideal
candidates for the bulk photovoltaic effect (BPVE) for visible wavelengths.
In this work we find strong nonlinear optical response in chiral crystal
insulators, which offers another interesting phenomena from the application
point of view.

The utilization of solar energy has been considered as one of the most convincing
strategies for future sustainable energy. To date, converting light to electricity via
a p-n junction-based solar cell is the most powerful technology for solar energy storage.
The output of the solar cell is constrained by the detailed balance limit, the probability
to photoexcite an electron to a conduction band is balanced by its decay back to a valence
band~\cite{shockley1961}. Hence, searching for new mechanisms to convert the light to electricity
is an important and necessary topic. Steady BPVE based on second
order optical response is one of the most promising candidates, where the intrinsic photocurrent
is generated by the polarized light in non-centrosymmetric
materials~\cite{kraut1979,belinicher1980,kristoffel1980,von1981,kristoffel1982,presting1982,aversa1995,sipe2000,yao2005,ma2017direct,xu2018electrically,morimoto2018current}.
The power conversion rate scales quadratically with the BPVE conductivity,
further the BPVE is determined by the intrinsic electronic band structure of a single crystal,
and a photovoltage above the band gap can be achieved via BPVE~\cite{yao2005,ji2010}, it is a
potential way to overcome the extrinsic limitations for solar power conversion.

Limited by the mechanism of the second order response, the size of the BPVE generated current is generally
much smaller than that from conventional p-n junction photovoltaics~\cite{young2012BFO,zheng2014,brehm2014}.
The shift current for the well-known Si-based 
solar cells is approximately
250 $\mu A/V^{2}$~\cite{pagliaro2008,rangel2017}, two orders of magnitude larger than the reported BPVE
in the typical ferroelectric materials of BaFeO$_3$~\cite{young2012BFO,zheng2014},
BaTiO$_3$~\cite{von1950,shieh2009}, PbTiO$_3$~\cite{young2012BFO}, etc. Very recently, a
design strategy was proposed for the large shift current via tuning the
tight binding (TB) model parameters, from which an effective 3D shift current of approximately 100 $\mu A/V^{2}$
was proposed~\cite{cook2017,rangel2017}.
However, a real material that satisfies the crucial TB parameters is still not realised. Therefore, real materials with
large shift current are desired for both fundamental study and potential applications.

In this work, we have found a group of insulators with chiral lattices
where the photoconductivity ranges from $\sim20$ to $\sim80$ $\mu A/V^{2}$,
and one order of magnitude larger than those reported in other materials.
This class of materials is defined by a chiral lattice structure that
cannot be obtained by a small interpolation between  centrosymmetric
and non-centrosymmetric structures.

\begin{figure}[htbp]
\begin{center}
\includegraphics[width=0.48\textwidth]{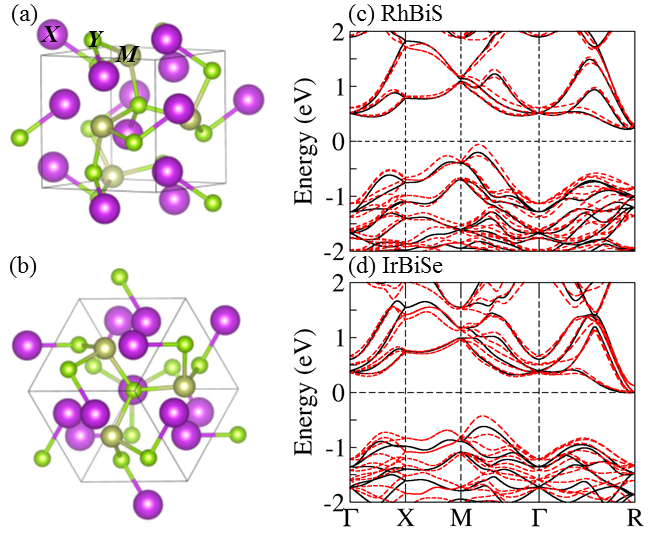}
\end{center}
\caption{
Representative crystal lattice and electronic band structures for the class of chiral
materials discussed in the main text.
(a, b) Crystal structure of this class of compounds $MXY$ ($M$=Ni, Pd, Pt, Rh, and Ir; $X$=P, As, Sb,
and Bi; $Y$=S, Se, and Te.) viewed from different directions.
(c, d) Energy dispersions along high symmetry lines for RhBiS and IrBiSe, respectively.
The bands without and with the inclusion of spin orbital coupling (SOC) are represented by a solid black line
and dashed red line, respectively.
}
\label{band}
\end{figure}

This group of compounds were first synthesized by F. Hulliger in 1963, with a cobaltite structure belonging
to space group $P213$(No.198), with a cubic setting, as presented in Fig.\ 1 (a,b)~\cite{hulliger1963}.
By substituting the three atoms in CoAsS by different elements, a class of new compounds $MXY$
(with $M$=Ni, Pd, Pt, Rh, and Ir; $X$=P, As, Sb, and Bi; $Y$=S, Se, and Te.) with the same crystal symmetry were synthesized.
With the absence of an inversion centre, they allow for a non-zero shift current. By considering the different electron
filling, one can further classify them into two subgroups of insulators ($M$=Rh and Ir) and metals ($M$=Ni, Pd, and Pt).

The crystal structure in Fig.1 (a,b) shows that only the transition metal ($M$) sites  can be tuned to inversion
positions by a small shift ($\sim3.5\%$ of lattice constants). For the sites $X$ and $Y$, all the atoms are
far away from the inversion sites. The atoms need to be moved by $\sim50\%$ of the lattice constants for at least four
atoms (two $X$ and two $Y$) to preserve the inversion symmetry. Therefore, the crystal lattice structure in this class
of compounds is far away from inversion symmetry.

Furthermore, the atoms at the $X$ and $Y$ sites are nearly in the
ionic form with strong local charge and spin orbital coupling (SOC). Therefore,
both giant spin-orbit splitting and large shift currents are expected.
Very recently, a giant Dresselhaus spin-orbit splitting at the top of the valence bands
was observed in one of these compounds, IrBiSe, via a good agreement between ARPES measurements and density functional
theory (DFT) calculations. This implies the accuracy of the first principles in this class of compounds~\cite{Liu2017IrBiSe}.

The shift current $J_{shift}(\omega)$ is generated by an electrical field
$\vec{E}(t)=\vec{E}(\omega)e^{i\omega t}+\vec{E}(-\omega)e^{-i\omega t}$ with photon energy
$\hbar\omega$ via the second order response
$J_{shift}^{c}(\omega)=\sigma_{ab}^{c}(\omega)E_{a}(\omega)E_{b}^{*}(\omega)$~\cite{kraut1979,kristoffel1980,von1981,sipe2000second}.
The photoconductivity $\sigma_{ab}^{c}$ is calculated by the quadratic response theory in the clean limit ~\cite{kraut1979,kristoffel1980,von1981,zhang2018photogalvanic},
\begin{equation}
\begin{aligned}
\sigma_{ab}^{c}=\frac{\left|e\right|^{3}}{8\pi^{3}\omega^{2}}Re\left(\underset{E_{\omega}=\pm\hbar\omega}{\sum}\underset{l,m,n}{\sum}\int_{BZ}d^{3}k(f_{l}-f_{n})\Omega_{l,m,n}^{abc}(\omega)\right)
\end{aligned}
\label{shift_current}
\end{equation}

\begin{equation}
\begin{aligned}
\Omega_{l,m,n}^{abc}(\omega)=\frac{<n|\hat{v}_{a}|l><l|\hat{v}_{b}|m><m|\hat{v}_{c}|n>}{(E_{n}-E_{m}-i\delta)(E_{n}-E_{l}+E_{\omega}-i\delta)}
\end{aligned}
\label{shift_current}
\end{equation}
where $|n>$ is an eigenvector of the Hamiltonian with eigenvalue $E_{n}$ at point $\vec{k}$ and $\hat{v}_{a}=\frac{\hat{p}_{a}}{m_{0}}$
is the velocity operator. To calculate the second order photoconductivity, we projected the
$ab-initio$ DFT Bloch wave function into atomic-orbital-like Wannier functions~\cite{first_wannier}
as performed in the code of the full-potential local-orbital
minimum-basis (FPLO)~\cite{Koepernik1999, perdew1996}.
Based on the highly symmetric Wannier functions, we constructed an effective tight-binding model Hamiltonian and calculated the photoconductivity
by following Formula (1).

Because all these compounds share the same symmetry operations and similar electronic band
structures, we consider only two cases as examples for detailed analysis.
One of them is RhBiS with the largest shift current in this family of materials,
and the other is IrBiSe with already confirmed band structure by recent ARPES
measurement~\cite{Liu2017IrBiSe}. The energy dispersions of RhBiS and IrBiSe
are shown in Fig.\ 1(c) and (d),
respectively. Both of them are semiconductors with band gaps around
0.5 eV. Because the top of the occupied bands are dominated by the Bi-6$p$ orbital,
SOC leads to a large spin splitting on the top of the valence bands, owing to the
absence of an inversion centre, which is fully consistent with recent ARPES measurements.
In addition, these classes of compounds host interesting band structures of triple point
fermions protected by the $c_3$ rotation symmetry and Weyl points with opposite chiralities
at different Fermi levels because of the absence of mirror symmetry.
These side products of our work will be discussed in detail in future independent reports.

In this compounds, the indirect band gap with SOC is smaller compared to the case without SOC
due to the spin split (see the dispersion in the $\Gamma-M$ direction), and the size of the decrease
depends on the strength of SOC. In addition, a direct result of the SOC spin splitting is the decrease
in the density of states (DOSs) near the top of the valence bands, from which we expect the decrease
in the joint DOS and the size of the photoconductivity for frequencies close to the band gap.

\begin{figure}[htbp]
\begin{center}
\includegraphics[width=0.5\textwidth]{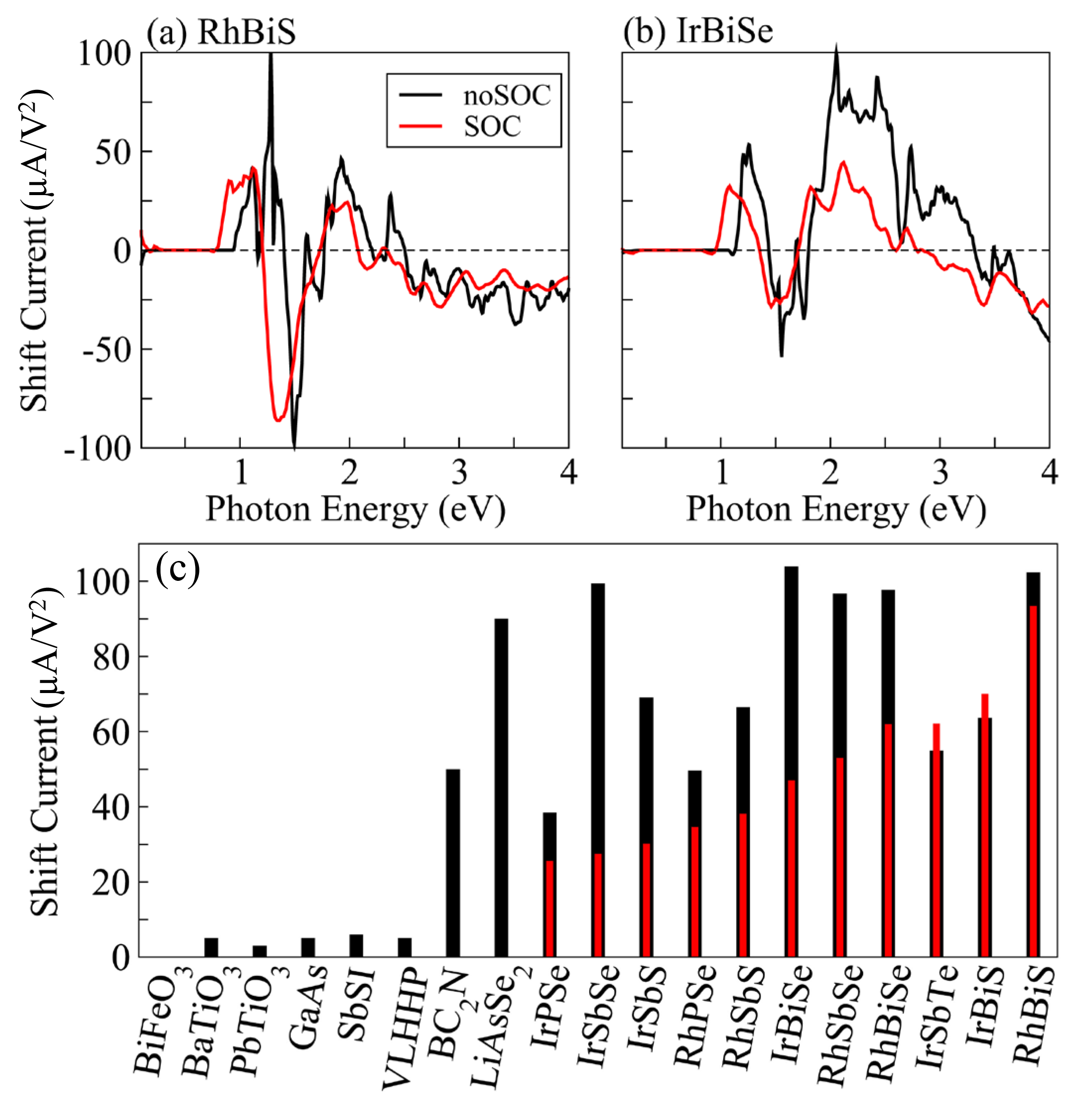}
\end{center}
\caption{
Three-dimensional shift currents for the insulators in the class of compounds $MXY$ considered here and other real materials reported previously.
(a) and (b) Energy-dependent shift current for RhBiS and IrBiSe, respectively.
(c) Comparison of the maximum values of shift currents with visible spectrum among the insulators obtained in this work and those reported in previous works.
The calculated results without and with the inclusion of SOC are presented by
black and red lines, respectively. The data for the shift currents are from \cite{young2012BFO,zheng2014,von1950,shieh2009,brehm2014first,cote2002rectification,sotome2019spectral}.
}
\label{insulator}
\end{figure}

Based on the second order response theory, $J_{shift}^{c}(\omega)=\sigma_{ab}^{c}(\omega)E_{a}(\omega)E_{b}^{*}(\omega)$,
the photoconductivity obtains a negative sign by inverting each of the indices. For the specified point group $T$, the three
$c_{2}$ symmetries with respect to the $x$, $y$, or $z$ axes play the crucial role of deciding the shape of the
photoconductivity tensor. For example, the rotation operation of $c_{2z}$ will change the sign of the tensor elements with
even times of the $z$-index, and therefore force them to be zero.
A similar rule also applies to $c_{2x}$ and $c_{2y}$.
Further, because of the $D_2$ subgroup, the three indexes give the same photoconductivity, and there is only one independent tensor element for the photoconductivity tensor in the shape of
\begin{equation}
\underline{\sigma}^{x}=\left(\begin{array}{ccc}
0 & 0 & 0\\
0 & 0 & \sigma\\
0 & \sigma & 0
\end{array}\right),
\underline{\sigma}^{y}=\left(\begin{array}{ccc}
0 & 0 & \sigma\\
0 & 0 & 0\\
\sigma & 0 & 0 \\
\end{array}\right),\underline{\sigma}^{z}=\left(\begin{array}{ccc}
0 & \sigma & 0\\
\sigma & 0 & 0\\
0 & 0 & 0
\end{array}\right)
\end{equation}

Our calculated results are fully consistent with the symmetry analysis.
Without considering SOC, RhBiS obtains the maximum value of $\sim-100$ $\mu A/V^{2}$
with photon energy $\sim 1.5$ eV, see Fig. 2(a). SOC shifts the whole curve of
the energy dependent photoconductivity to a lower frequencies by an amount $\sim0.3$ eV.
The peak is moved to a relatively small photon energy of $\sim 1.4$ eV, and the maximum value
is reduced to $\sim 80$ $\mu A/V^{2}$.
Similar behavior can be also found in IrBiSe, see Fig. 2(b). Without including SOC,
the photoconductivity presents two peak values
that can reach 100 $\mu A/V^{2}$ with a photon energy of $\sim 2.1$ and
$\sim 2.5$ eV, respectively. After taking SOC into consideration, the
photoconductivity reduces to only $\sim50$ $\mu A/V^{2}$.
This decrease is larger than that in RhBiS, possibly due to stronger SOC.
Hence, SOC plays a negative role for the shift current, as expected from band
structure analysis.

The comparison of photoconductivity among the compounds reported from the previous
literature and the semiconductors from this work is given in Fig.\ 2(c). Two typical
cases are the ferroelectrics BaTiO$_3$ and PbTiO$_3$, in which the photoconductivities are
$\sim5$ $\mu A/V^{2}$ in the photo energy in the visible range. So far, the largest
photoconductivity in the visible spectrum range is
$\sim 6$ $\mu A/V^{2}$, observed in SbSI~\cite{sotome2019spectral} in 2018. In the class of
semiconductors of the present work, the photoconductivity can range from
$\sim 20$ $\mu A/V^{2}$ to $\sim 80$ $\mu A/V^{2}$, around one order magnitude larger
than previous reports, a new record for BPVE in three-dimensional
compounds.

Normally, the decrease of photoconductivity by SOC only works in simple
band structures, where we assumed that the system contains a few bands only and
that the sub-bands after the SOC spin split do not mix with each other.
If the band structure is complicated and with strong hybridization, SOC
might change the band orders and contribute positively to the shift current.
SOC also complicates the hybridization of different bands for this class of
 compounds, see Fig. 1 (c-d). As shown in Fig. 2(c), the photoconductivity
decreases after the inclusion of SOC for most of the insulators in this class
of compounds. The only two exceptions are IrSbTe and IrBiS, in which the
competition of different effects from spin splitting almost cancel each other.

\begin{figure}[htbp]
\begin{center}
\includegraphics[width=0.48\textwidth]{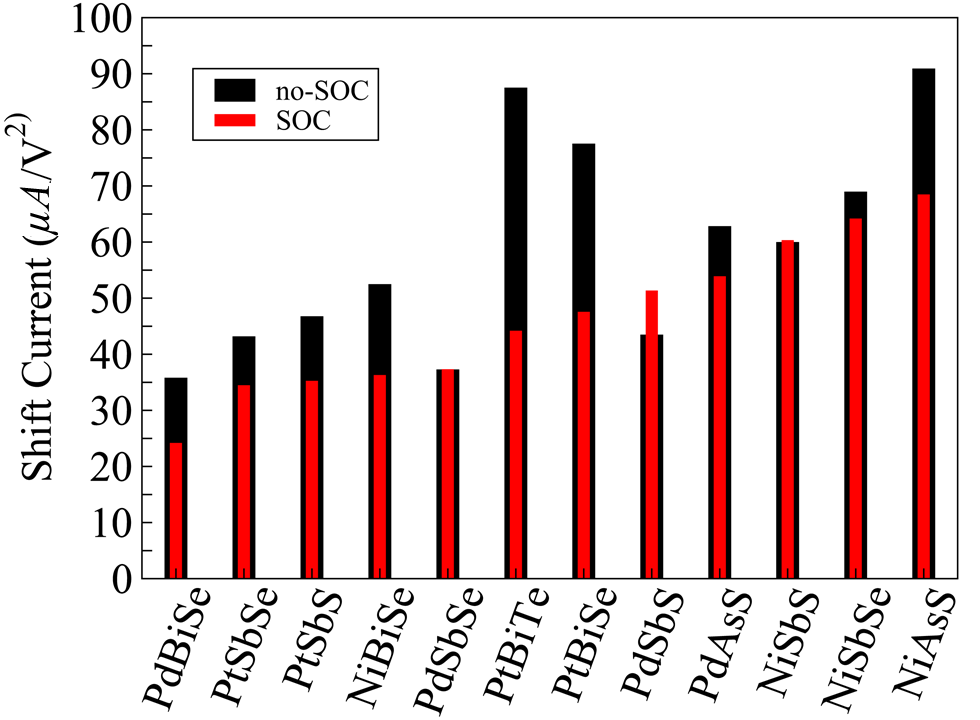}
\end{center}
\caption{
Peak value of three-dimensional shift currents in the visible spectrum of metals studied in this work.
The calculated results without and with the inclusion of SOC are presented by
black and red lines, respectively.
}
\label{metal}
\end{figure}

Because the DFT calculations with generalized gradient approximations (GGA) normally
underestimate the band gap in insulators~\cite{perdew1996}, we double-checked the
band structures by using a more accurate hybrid functional~\cite{heyd2003hybrid}.
It is found that the only difference in the band structures from the two types
of approximations is the band gap, and the shapes of the bands almost do not change
for both valence and conduction bands near the band gap, see Supplementary Fig.\ 1.
Hence, we believe that the maximum values of the photoconductivity should also shift
to higher frequencies, while keeping the magnitude almost unchanged.
The comparison between the band structure from the hybrid functional and GGA
calculations yields that the frequency shift is around 0.5 eV.

Very recently, considerable interest regarding BPVE was devoted to metallic systems,
as the natural requirement of inversion symmetry breaking for BPVE can also lead
to topological Weyl semimetals
(WSMs) with monopoles of Berry curvature and linear energy
dispersion~\cite{Gavin2018,cook2017,yang2018flexo,morimoto2016topological}.
The gapless electronic band structure with inversion symmetry broken provides the
possibility to extend BPVE into the long wavelength range. Indeed, strong shift
current can be also obtained in inversion-symmetry-broken
metals. In this class of compounds, there are 12 metals with a transition metal
atom sitting at $M$=Ni, Pd, and Pt, where the photoconductivity can range from
$\sim 30$ to $\sim 60$ $\mu A/V^{2}$ in the visible spectrum range, around one
order magnitude larger than that reported for ferroelectrics. Similar to the
insulators, SOC also plays a negative role most of the time, see Fig.\ 3.

In summary, we theoretically predicted strong shift currents in chiral crystal materials.
The three-dimensional photoconductivity can range from $\sim20$
to $\sim80$ $\mu A/V^{2}$, which is around one order of magnitude larger than for other reported real materials.
Via electronic band structure analysis, we found that SOC plays a negative role in most cases for the shift current.
This class of compounds provides a good platform to study the mechanism of the PBVE and its future applications.
It illustrates that chiral crystals can host not only quantized CPGE but also a strong shift current, of great technological importance.

\begin{acknowledgments}
This work was financially supported by the ERC Advanced Grant No.\ 291472 `Idea Heusler', ERC Advanced Grant No.\ 742068 `TOPMAT'.
This work was performed in part at the Center for Nanoscale Systems (CNS), a member of the National Nanotechnology Coordinated Infrastructure Network (NNCI), which is supported by the National Science Foundation under NSF award no. 1541959. CNS is part of Harvard University.
Some of our calculations were carried out on
the Cobra cluster of MPCDF, Max Planck society.

\end{acknowledgments}

\bibliography{topology}
\end{document}